\documentclass[%
 reprint,
 amsmath,amssymb,
 aps,prl,
]{revtex4-1}

\usepackage{graphicx}
\usepackage{dcolumn}
\usepackage{bm}

\begin{document}
\pdfoutput=1

\title{Significant reduction of electronic correlations upon isovalent Ru substitution of BaFe$_2$As$_2$ }%

\author{V. Brouet$^1$, F. Rullier-Albenque$^2$, M. Marsi$^1$, B. Mansart$^1$, M. Aichhorn $^3$, S. Biermann$^3$, J. Faure$^{4,5}$, L. Perfetti$^5$, A. Taleb-Ibrahimi$^6$, P. Le F\`evre$^6$, F. Bertran$^6$, A. Forget$^2$ and D. Colson$^2$}

\affiliation{$^1$Laboratoire Physique des Solides, Univ. Paris Sud, CNRS-UMR8502, 91405 Orsay, France\\
$^2$ Service de Physique de l'Etat Condens\'e,CEA Saclay,CNRS-URA2464,91191 Gif sur Yvette,France\\
$^3$ Centre de Physique Théorique, École Polytechnique, CNRS, 91128 Palaiseau, France\\
$^4$ Laboratoire d'Optique Appliqu\'ee,ENSTA,Ecole Polytechnique,CNRS-UMR7639,91761 Palaiseau,France\\ 
$^5$ Laboratoire des Solides Irradi\' es, Ecole Polytechnique, CNRS-CEA/DSM, 91128 Palaiseau,France\\
$^6$  Synchrotron SOLEIL,CNRS, Saint-Aubin-BP48,91192 Gif-sur-Yvette,France}


\begin{abstract}
We investigate Ba(Fe$_{0.65}$Ru$_{0.35}$)$_2$As$_2$, a compound where superconductivity appears at the expense of magnetism, by transport measurements and Angle Resolved photoemission spectroscopy. By resolving the different Fermi Surface pockets and deducing from their volumes the numbers of holes and electrons carriers, we show that Ru does neither induce hole nor electron doping. However, the Fermi Surface pockets are about twice larger than in BaFe$_2$As$_2$.  A change of sign of the Hall coefficient with decreasing temperature evidences the contribution of both carriers to the transport. Fermi velocities increase significantly with respect to BaFe$_2$As$_2$, suggesting a reduction of correlation effects. 
\begin{description}
\item[PACS numbers]
74.70.Xa, 74.25.Jb, 74.25.F-
\end{description}

\end{abstract}

\maketitle

A key similarity between iron pnictides (FePn) and cuprate superconductors is the proximity of magnetic and superconducting phases in their phase diagrams. BaFe$_2$As$_2$ is for example a semi-metal with the same number of holes and electrons carriers (in the following, we refer to this situation as ``undoped") and it undergoes a Spin Density Wave (SDW) transition below 139K \cite{RotterMag}. However, whereas superconductivity is only observed in doped cuprate superconductors, there is a number of ways to induce it in FePn. In the BaFe$_2$As$_2$ family, it can be obtained through hole \cite{Rotter} or electron \cite{Sefat} doping, but also by applying pressure \cite{Alireza} or through isovalent substitutions, either P at the As site \cite{Jiang} or Ru at the Fe site \cite{Paulraj}. It is likely that, whereas in cuprate superconductors the Mott insulating state has to be destroyed by doping before superconductivity may emerge, the itinerant nature of magnetism in BaFe$_2$As$_2$ makes it easier to switch to superconductivity. In this respect, FePn may be closer to heavy fermions systems \cite{Mathur}, although they have much higher superconducting temperatures. Obviously, understanding such evolutions should give crucial information about the role of correlations and magnetism in the formation of the superconducting state.

Distressingly, very little information is available so far about the changes of the electronic structure leading to superconductivity in undoped compounds. It is not even established whether substitutions or pressure do or do not induce an effective doping. Particularly in the case of Ru/Fe substitutions, the possible formation of Ru$^{4+}$ and Ru$^{5+}$ (instead of a valence 2+ for Fe in the FeAs slab) could result in electron doping \cite{Zhang}. The first study of these compounds \cite{Paulraj} in fact concluded that these systems were electron doped. On the contrary, a density functional theory calculation \cite{Zhang} predicts these systems to behave as coherent alloys, with an increased bandwidth, due to the larger spatial extension of the Ru 4d orbitals. This larger bandwidth would lower the density of states at the Fermi level n(E$_F$) and this may be the reason for the destabilization of the magnetic state. Similarly, changes of the structure under pressure or P/As substitution are under scrutiny to explain the appearance of superconductivity. Especially, it is known that the band structure is extremely sensitive to the Fe-As-Fe bond angle \cite{Vildosola}, which changes in all these different cases \cite{Kimber,Zhao}. Clearly, direct experimental determination of the band structure in undoped superconductors is highly desirable to go further.  

We present here a detailed investigation of Ba(Fe$_{0.65}$Ru$_{0.35}$)$_2$As$_2$ (abbreviated in the following as BaFeRu$_{0.35}$As), a compound that is non-magnetic and superconducting at 19.5K. In contrast to BaFe$_2$As$_2$, where the Hall coefficient R$_H$ is always negative \cite{Rullier}, R$_H$ becomes positive here below 115K, indicating a larger mobility of holes in BaFeRu$_{0.35}$As. By Angle Resolved Photoemission Spectroscopy (ARPES), we detect well defined hole and electron bands, evidencing an homogeneous electronic structure and no strong disorder associated with Ru substitutions. The structure and nesting properties of the Fermi Surface (FS) are similar to that of other compounds of this family \cite{Ding,Liu,YiShen2,Terashima,Brouet,Vilmercati,Malaeb,Thirupathaiah}, suggesting identical orbital origin for the different pockets. However, the electronic structure is strongly modified compared to BaFe$_2$As$_2$ : The number of both carriers (holes and electrons) has doubled and the Fermi velocities (v$_F$) have increased by a factor nearly 3. No similar tendency was observed under hole or electron doping, for which a simple rigid band filling picture applies well \cite{Ding,Brouet}. These changes largely exceed those expected in band structure calculations we performed for BaFe$_2$As$_2$ and BaFeRuAs$_2$. We conclude that there is a substantial reduction of the bands renormalization upon Ru substitution, hence of correlation effects. The consequences of this situation for the competition between magnetism and superconductivity will be interesting to explore. 

\begin{figure}[t]
\includegraphics[width=7cm]{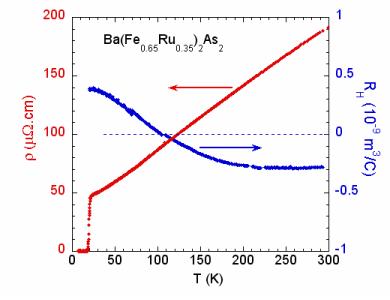}
\caption{\label{fig:Transport} Temperature dependence of the in-plane resistivity (left) and of the Hall coefficient R$_H$ (right) in BaFeRu$_{0.35}$As.}
\end{figure}

We have synthesized single crystals of Ba(Fe$_{1-x}$Ru$_x$)$_2$As$_2$ with sizes up to 0.5*0.4*0.02~mm$^3$, using a self-flux method \cite{Rullier2}. We find that the a lattice parameter increases slightly with x, whereas c decreases, in agreement with ref. \cite{Paulraj}. For x=0.35, a = 4.0344(5)\AA~and c = 12.760(3)\AA, which means that the Fe-As-Fe bond angle increases from 111.18$^{\circ}$ to 113.73$^{\circ}$. Fig.~\ref{fig:Transport} shows that, at x=0.35, corresponding approximately to optimal doping \cite{Rullier2}, the resistivity does not exhibit any sign of the SDW transition anymore and a superconducting state is stabilized under 19.5K. Moreover, Hall effect measurements performed in the Van der Pauw configuration in magnetic fields up to 14T show that the Hall coefficient, which is negative at high temperature becomes positive below 115K. This behavior is very different from that observed in undoped and Co-doped BaFe$_2$As$_2$, where R$_H$ just decreases with temperature \cite{Rullier}. It is also different from that reported previously for polycrystalline samples of Ba(Fe$_{1-x}$Ru$_x$)$_2$As$_2$ \cite{Paulraj}, where R$_H$ was found negative. This was interpreted as a sign of electron doping induced by Ru, but the present result in single crystals casts doubts on this conclusion. In a two band model, our positive R$_H$ at low temperatures indicates that the mobility of holes overcomes the one of electrons in this limit. This implies that some scattering processes responsible for the low mobility of holes in BaFe$_2$As$_2$ are considerably reduced by the introduction of Ru. A complete analysis of the transport properties as a function of Ru doping is reported elsewhere \cite{Rullier2}. 

\begin{figure}[t]
\includegraphics[width=8cm]{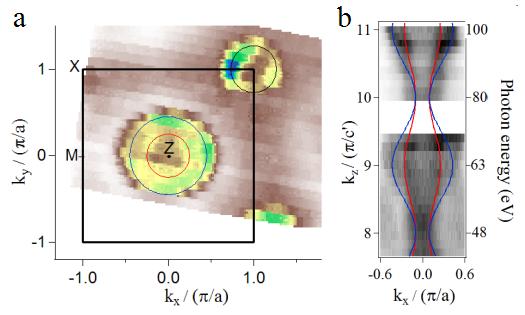}
\caption{\label{fig:FS} (a) Fermi Surface of BaFeRu$_{0.35}$As at 20K, with 100eV photon energy (i.e. k$_z$=11$\pi$/c') and LHP. Spectral weight is integrated in a 10meV window. Black square is the BZ of the 3D unit cell. Circles indicate the hole (red and blue) and electron (black) pockets, as determined in the text. (b) Spectral intensity at E$_F$ measured along k$_x$ as a function of photon energy (right) or, equivalently, of k$_z$ (left). Blue and red lines are cosine dispersion with k$_z$ periodicity for the two hole pockets (see supplementary information. }
\end{figure}

\begin{figure}[t]
\includegraphics[width=7.5cm]{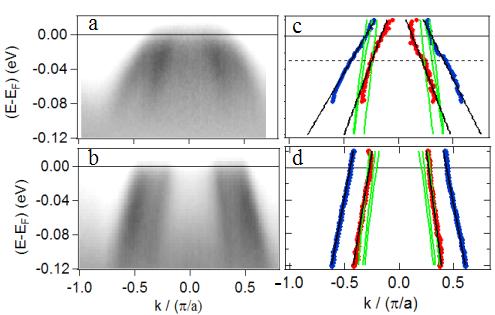}
\caption{\label{fig:Holes} Hole pockets : (a) ARPES intensity plot measured around Z in BaFe$_2$As$_2$, at 150K, with 34eV photon energy (k$_z$=7$\pi$/c') and LHP. (b) Same for BaFeRu$_{0.35}$As at 20K, with 65eV photon energy (k$_z$=9$\pi$/c') and LHP. (c) and (d) Points : Dispersions extracted by a 4 lorentzian fit of the Momentum Distribution Curves (MDC) of the left images, with linear fits. Green lines are the calculated dispersions along ZM for the three hole bands (see text).}
\end{figure}

\begin{figure}[t]
\includegraphics[width=8.5cm]{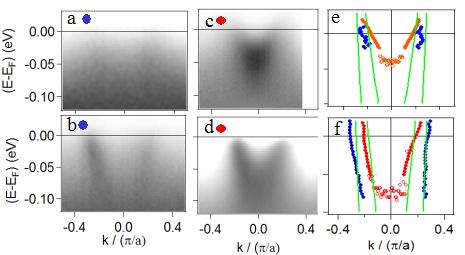}
\caption{\label{fig:Electrons} Electron pockets (the center was taken as k=0) : (a) ARPES intensity plot measured in BaFe$_2$As$_2$ at 150K, with 92eV photon energy and LHP, (c) same with 34eV and LVP. (b) Same in BaFeRu$_{0.35}$As at 20K, with 100eV photon energy and LHP, (d) same with 45eV and LVP. (e) and (f) Dispersions of the left images extracted by a MDC fit. Bottom of the band (open circles) is defined as maximum of the EDC, when possible. Green lines are dispersions obtained by band structure calculations (see text) for the two electron bands along $\Gamma$X.}
\end{figure}

Investigation of the electronic structure by ARPES clarifies changes in the electronic structure. Experiments were carried out at the CASSIOPEE beamline of the SOLEIL synchrotron. The photon energy was tuned between 35 and 100eV and the polarization of the light was linear, either in the plane of incidence (named hereafter LHP for Linear Horizontal Polarization) or perpendicularly to it (Linear Vertical Polarization, LVP). Fig.~\ref{fig:FS}a presents the FS measured at 20K and k$_z$=$\pi$/c' (k$_z$ is the direction perpendicular to the FeAs slab, it is quoted modulo 2$\pi$/c', where c'=c/2 is the distance between two FeAs slabs. It is fixed by the photon energy h$\nu$, through k$_z$=0.512$\sqrt{h\nu-W+V_0}$, with V$_0$$\approx$14~eV in these compounds, see ref.\cite{Brouet,Vilmercati}). The FS structure is basically similar to that of BaFe$_2$As$_2$, either pure or substituted with K (hole doping) or Co (electron doping) \cite{Ding,Brouet,Liu,YiShen}. Two different circular hole pockets are present at the Brillouin Zone (BZ) center Z (called hereafter $\alpha$ (inner red circle) and $\beta$ (outer blue circle), the sizes of the circles were determined from Fig.~\ref{fig:Holes}). Doubly degenerate electron pockets are found at the BZ corners X (black circle, see also Fig.~\ref{fig:Electrons}). Fig.~\ref{fig:FS}b further shows that, as in BaFe$_2$As$_2$ \cite{Brouet,Vilmercati,Malaeb}, the sizes of the hole pockets appear strongly dependent on the photon energy, indicating substantial k$_z$ dispersion (see supplementary information for more details). The similar ratio and photon energy dependence of the two hole bands suggest that they are formed by the same orbitals, possibly with slightly stronger 3D effects in  BaFeRu$_{0.35}$As. On the contrary, the electron pockets do not exhibit strong k$_z$ dependence, although they are quite sensitive to experimental conditions (see Fig.~\ref{fig:Electrons}). This is again similar to other FePn \cite{Brouet,Malaeb}. Note that, at first sight, the electron pockets would nest rather well into the inner hole pocket, if translated by ($\pi$/a, $\pi$/a). The suppression of the SDW state then cannot be simply attributed to a worse FS nesting. Rather good nesting was also found in other FePn superconductors \cite{Ding,Terashima}.  

A major difference between these FS, however, is that the sizes of the hole pockets are much larger in BaFeRu$_{0.35}$As than in BaFe$_2$As$_2$. As the volume of the pockets are simply proportional to the number of holes or electrons they contain, this is an important information to consider. The diameter of the hole pockets can be accurately determined by the crossing of the hole bands with E$_F$. At k$_z$=$\pi$/c', Fig.~\ref{fig:Holes} indicates k$_F$$^{\alpha}$=0.25$\pi$/a and k$_F$$^{\beta}$=0.42$\pi$/a for BaFeRu$_{0.35}$As. This is almost twice larger than in BaFe$_2$As$_2$ (k$_F$$^{\alpha}$=0.14$\pi$/a and k$_F$$^{\beta}$=0.28$\pi$/a) and rather resembles the hole-doped Ba$_{0.6}$K$_{0.4}$Fe$_2$As$_2$ (k$_F$$^{\alpha}$= 0.22$\pi$/a and k$_F$$^{\beta}$= 0.43$\pi$/a \cite{Ding}). Integrating the volume of the hole pockets over k$_z$ using the contours shown in Fig. 2b (see supplementary information for details), we estimate they contain n$_h$=0.11 holes/Fe. If BaFeRu$_{0.35}$As is undoped, one expects n$_h$=n$_{el}$, hence k$_F$$^{el}$~=0.26$\pi$/a for two circular and degenerate electron pockets. This is indicated as a black circle in Fig.~\ref{fig:FS}a and agrees very well with the actual size of the electron pocket. \textit{We therefore conclude that BaFeRu$_{0.35}$As is a compensated semi-metal with n=n$_h$=n$_{el}$= 0.11 carriers/Fe}. 

This number of carriers is significantly larger than the value we estimated in BaFe$_2$As$_2$ n=0.06$\pm$0.02 with a similar reasoning \cite{Brouet}. This leads to the very different contributions of holes and electrons in the transport properties of these two compounds observed in Fig.~1 \cite{Rullier2}. Ortenzi \textit{et al.} proposed that a reduced number of carriers compared to theoretical estimates (n=0.15 \cite{XuFang}) could be due to interband interactions \cite{Ortenzi}. This may then be a sign of different band interactions and we will return to this point later.   

The strong difference between the two electronic structures is further revealed by the comparison of their band dispersions. Fig.~\ref{fig:Holes} compares the dispersions at photon energies equivalent to k$_z$=$\pi$/c', where the bands are most clearly separated (data in BaFe$_2$As$_2$ are shown at 150K to avoid complications due to the magnetic phase). In BaFeRu$_{0.35}$As, the dispersions are clearly much steeper, they are almost parallel and a linear fit yields v$_F$$^{\alpha}$=1.16eV.\AA~ and v$_F$$^{\beta}$=0.89 eV.\AA. In BaFe$_2$As$_2$, the bands are more rounded and a linear fit near the Fermi level yields v$_F$$^{\alpha}$=0.43eV.\AA~ and v$_F$$^{\beta}$=0.32eV.\AA, a factor 2.75 smaller. Although v$_F$ formally depends on k$_F$ (v$_F$=$\hbar$k$_F$/m$^*$), this difference is not due to the different k$_F$. The slope of the dispersions in BaFe$_2$As$_2$ would be nearly the same if measured at -0.03eV (dotted line in Fig.~\ref{fig:Holes}b), where the crossings occurs at values close to k$_F$ in BaFeRu$_{0.35}$As. Also, in Ba$_{0.6}$K$_{0.4}$Fe$_2$As$_2$, where there is almost the same number of holes as in BaFeRu$_{0.35}$As, the dispersions are still very different, v$_F$$^{\alpha}$=0.5eV.\AA~and v$_F$$^{\beta}$=0.22eV.\AA~\cite{Ding}. Therefore, there is a strong intrinsic increase of v$_F$ in BaFeRu$_{0.35}$As.

To understand whether this increase is due mainly to changes in the band structure or to a different renormalization of the band structure, we performed electronic structure calculations within the local density approximation, using the Wien2K package \cite{Wien2k}. For BaFe$_2$As$_2$, we considered the tetragonal structure with cell parameters taken from \cite{RotterMag}. The substitution by Ru has been studied for BaFeRuAs$_2$, where one of the two Fe atoms in the unit cell has been replaced by Ru, and with cell parameters reported here for BaFeRu$_{0.35}$As. Although this composition is slightly larger than the one studied here (x=0.5 instead of x=0.35), this gives a first idea of the qualitative evolution. In Fig.~\ref{fig:Holes}c and Fig.~\ref{fig:Holes}d, these calculations are compared to the dispersions determined experimentally. For BaFe$_2$As$_2$, a renormalization by a factor about 3 would be needed to align the dispersions, in agreement with previous studies \cite{YiShen} and theoretical expectations of correlation strength in FePn \cite{Aichhorn}. On the contrary, it would be neglegible for BaFeRu$_{0.35}$As. \textit{The increase of v$_F$ is then not primarily driven by changes in the band structure, but by significantly smaller correlation effects}. 

Fig.~\ref{fig:Holes} further shows that the three hole bands should be shifted with respect to E$_F$ to obtain a good agreement with experiment. These shifts should mostly enlarge the pockets in BaFeRu$_{0.35}$As and shrink them in BaFe$_2$As$_2$. This corresponds to the larger FS in the former compound we have already discussed, but further indicates that this change is not expected at the band structure level and should then be associated to correlations and/or different interband interactions \cite{Ortenzi}. These shifts are likely orbital sensitive and may also depend on k$_z$, so that a thorough comparison between theory and experiment, including orbital symmetries, 3D effects, exact dopings and structures, will be very important to fully quantify the nature of these changes. 

The behavior of the electron pockets presented in Fig.~\ref{fig:Electrons} is consistent with that of the holes. Band structure calculations (Fig.~\ref{fig:Electrons}e and \ref{fig:Electrons}f) predict two electron bands, with quite different v$_F$ for the inner and outer bands. These two bands are usually not clearly resolved experimentally, but we observed empirically that the contribution of the outer band is dominant for photon energy around 100eV and LHP (Fig.~\ref{fig:Electrons}a and \ref{fig:Electrons}b) and that of the inner band for lower photon energies and LVP (Fig.~\ref{fig:Electrons}c and \ref{fig:Electrons}d). Because the bands cannot be clearly separated, it is more difficult to compare v$_F$ directly \cite{asymmetry}. Nevertheless, the tendency of smaller renormalization for BaFeRu$_{0.35}$As holds. We also observe that the bottom of the inner electron band seems to shift from 40meV to 70meV in BaFeRu$_{0.35}$As (open circles in Fig.~\ref{fig:Electrons}c and \ref{fig:Electrons}f), suggesting larger bandwidth, consistent with the larger Fermi velocities. Better separation of the two bands will be needed for more quantitative comparison.  

Finally, we have established that the electronic structure is qualitatively similar in BaFeRu$_{0.35}$As and BaFe$_2$As$_2$, as far as the number of holes and electrons pockets, their 3D character and probably their main orbital origin is concerned. However, correlation effects appear strongly reduced in BaFeRu$_{0.35}$As and, in fact, almost negligible compared to our LDA calculations. This can explain the increased mobility of holes in BaFeRu$_{0.35}$As compared to BaFe$_2$As$_2$ revealed by Hall measurements. The band width we calculate in BaFeRuAs$_2$ is substantially larger ($\sim$6eV) than that of BaFe$_2$As$_2$ ($\sim$4eV), similarly to \cite{Zhang}, naturally leading to weaker correlations, even for comparable interactions. We attribute this to increased covalency effects of the bigger Ru ions compared to Fe. These different correlations strengths may have important consequences for the ordering of the different orbitals and finely tune the orbital weights at the FS. As the occupation of the different orbitals probably depends sensitively on the correlations, it is worth stressing that we simultaneously observe quite a different number of carriers compared to BaFe$_2$As$_2$. The resulting electronic structure of this undoped FePn superconductor is then markedly different from that of doped FePn superconductors, despite their similar T$_c$. Interestingly, a de Haas-Van Alphen investigation in BaFe$_2$(As$_{1-x}$P$_x$)$_2$ recently reported larger FS areas and lower effective mass upon P substitution \cite{Shishido}. This suggests that the two effects are really linked and that this modification is characteristic of undoped superconductors. The delicate balance between the different orbitals may play a crucial role for stabilization of magnetism or superconductivity. Furthermore, this example shows that one can tune quite significantly the band structure and correlations in iron pnictides, revealing notably that the absolute number of carriers is probably an additional important parameter. 
 
We thank H. Alloul, J. Bobroff, Y. Laplace and A. georges for discussions and P. Thu\'ery for help with structural measurements. M.A. acknowledges financial support from the Austrian Science Fund, project J2760.


\begin{thebibliography}{29}
\bibitem{RotterMag} M. Rotter et al., Phys. Rev. B 78, 020503(R) (2008)
\bibitem{Rotter} M. Rotter, M. Tegel and D. Johrendt, Phys. Rev. Lett 101, 107006 (2008)
\bibitem{Sefat} A.S. Sefat et al., Phys. Rev. Lett. 101, 117004 (2008)
\bibitem{Alireza} P.L. Alireza et al., J. Phys. Cond. Mat 21, 012208 (2009)
\bibitem{Jiang} S. Jiang et al., J. Phys. Cond. Mat. 21, 382203 (2009)
\bibitem{Paulraj} S. Sharma et al., Phys. Rev. B 81, 174512 (2010)
\bibitem{Mathur} N.D. Mathur et al., Nature 394, 39 (1998)
\bibitem{Zhang} L. Zhang and D.J. Singh, Phys. Rev. B 79, 174530 (2000)
\bibitem{Vildosola} V. Vildosola et al., Phys. Rev. B 78, 064518 (2008)
\bibitem{Kimber} S.A.J. Kimber et al., Nature 8, 471 (2009)
\bibitem{Zhao} J. Zhao et al., Nature Materials 7, 953 (2008)
\bibitem{Rullier} F. Rullier-Albenque and D. Colson and A. Forget and H. Alloul , Phys. Rev. Lett. 103, 057001 (2009)
\bibitem{Ding}H. Ding et al., EPL 83, 47001 (2008); H. Ding et al., cond-mat/0812.0534; Y. Xu et al., cond-mat/0905.4467
\bibitem{Liu} C. Liu et al., Phys. Rev. Lett. 101, 177005 (2008)
\bibitem{YiShen2} M. Yi et al., Phys. Rev. B 80, 174510 (2009)
\bibitem{Terashima} K. Terashima et al., PNAS 106, 7330 (2009)
\bibitem{Brouet} V. Brouet et al., Phys. Rev. B 80, 165115 (2009)
\bibitem {Vilmercati} P. Vilmercati et al., Phys. Rev. B 79, 220503(R) (2009)
\bibitem{Malaeb} W. Malaeb et al., J. of the Phys. Soc. of Japan 78, 123706 (2009)
\bibitem{Thirupathaiah} S. Thirupathaiah et al., Phys. Rev. B 81, 104512 (2010)
\bibitem{Rullier2} F. Rullier-Albenque et al., Phys. Rev. B 81, 224503 (2010)
\bibitem{YiShen} M. Yi et al., Phys. Rev. B 80, 024515 (2009)
\bibitem{XuFang} G. Xu and H. Zhang and X. Du and Z. Fang, EPL 84, 67015 (2008)
\bibitem{Ortenzi} L. Ortenzi and E. Capelluti and L. Benfatto and L. Pietronero, Phys. Rev. Lett. 103, 046404 (2009)
\bibitem{Wien2k} P. Blaha et al., Wien2K, ISBN 3-9501031-1-2 (2002)
\bibitem{Aichhorn} M. Aichhorn et al., Phys. Rev. B 80, 085101 (2008)
\bibitem{asymmetry} The asymmetry of the dispersion sometimes observed between the left and right sides of the electron pockets is also likely due to a different combination of the two electron bands on the two sides.
\bibitem{Shishido} H. Shishido et al., Phys. Rev. Lett 104, 057008 (2010)
\end{thebibliography}

\bigskip

\bigskip
\bigskip
{\bf Supplementary information}
	
\begin{figure}[h]
\includegraphics[width=8.5cm]{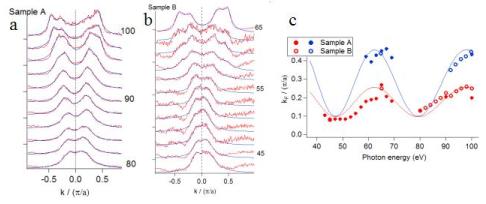}
\caption{\label{fig:3Defects} 
(a) and (b) MDC taken at E$_F$ as a function of the indicated photon energies, for two different samples of BaFeRu$_{0.35}$As. They are fitted by 4 lorentzians, when the $\alpha$ and $\beta$ bands can be clearly resolved, by 2 lorentzians otherwise. The positions of these lorentzians are reported in (c) as a function of photon energy. }
\end{figure}

	Both bands clearly shift with photon energies, which indicates sizable 3D effects. However, it becomes difficult to resolve 3D effects when the bands overlap, especially as their relative intensity may change with photon energy. It is for example difficult to decide whether $\beta$ merges with $\alpha$ near k$_z$=0 or disappears, due to accidental suppression by matrix element effects. Similarly, the slightly different MDCs observed at equivalent k$_z$=0 points (e.g. 45eV and 80eV, a difference that is reproducible in the two samples) is best explained by a two peak structure with different relative intensity at these two photon energies. The two peaks could be a and b, but also two bands forming a (remember three hole pockets are expected in the calculation).

	This situation is qualitatively very similar to that of pure and Co-substituted BaFe$_2$As$_2$, where the diameters of the hole pockets also appear strongly reduced near k$_z$=0 [17-20]. This suggests that 3D effects are similar in all these compounds, although their magnitude may be slightly different in the different cases (this was suggested in ref. [20] for Co substitutions). 

	If the holes FS sheets were cylinders, the number of holes per Fe would be given by n$_h$=$\pi$(k$_F$$^{\alpha}$)$^2$/2+$ \pi$(k$_F$$^{\beta}$)$^2$/4, assuming $\alpha$ is doubly degenerate and $\beta$ is singly degenerate [17]. However, these sheets are clearly warped along k$_z$, so that a model of the dispersion along k$_z$ is needed to correctly evaluate the volume of the pockets. Assuming the cosine dispersions shown as red and blue lines in the figure above and Fig. 2b, which seems the most natural choice to us, we obtain n$_h$=0.11 holes/Fe. Note that if the k$_z$ dispersion of $\beta$ as smaller, the number of holes would slightly increase. One would for example get n$_h$=0.14 holes/Fe for two bands with average diameters k$_F$=0.15$\pi$/a and k$_F$=0.37$\pi$/a.

            In ref. [17], we used a similar model of the hole pockets dispersion and degeneracy to estimate n$_h$=0.06+/-0.02 in BaFe$_2$As$_2$. The relative difference in the number of holes will remain the same, if other choices were made to describe 3D effects, unless their shape changes between the two compounds. This could reduce the difference, but the number of holes will remain larger in BaFeRu$_{0.35}$As, as the pocket is already somewhat larger at k$_z$=0.

\end{document}